\documentclass [10pt,a4paper]{article}
\usepackage{amssymb}
\begin{document}

\title{Modelling  comonotonic group-life
under dependent  decrement causes}
\author{ Dabuxilatu Wang \footnote{The research has been supported by
the scientific foundation 2006, No.62033 of Guangzhou Education
Bureau.}\\ School of mathematics and information sciences,\\
Guangzhou University,
 No.230 Waihuan Xilu, \\ Higher Education Mega Center, Guangzhou, 510006,
 PR China.}
\date{}
\maketitle
\begin{abstract}
Comonotonicity had been a extreme case of dependency between
random variables.  This article consider an extension of single
life model under multiple dependent decrement causes  to the case
of comonotonic group-life.

\end{abstract}

\section{Introduction}
The concept of comonotonicity was proposed for research of
economics,  and it had been introduced into the insurance
actuarial in 1990's (see. Dhaene J. {\em et al.}\cite{dh}) for
ordering risks and stop-loss premium determination (see. Dhaene J.
{\em et al.} \cite{dh1}). Research results show that
comonotonicity had been a extreme case of dependency between
random variables (see. Dhaene J. {\em et
al.}\cite{dh},\cite{dh1},\cite{dh2}, Ribas C.\cite{ri}). Thus, for
example, comonotonicity between lives in a group implies a special
characteristics of the group-life that one member of the group
attains venerable age then all of other members of the group  also
attain venerable age. The decrement models with respect to multiple
dependent causes of withdrawal have been received a  attention
from life insurance area, Carriere JF.(1994,1998)
\cite{ca},\cite{ca1} and Vladimir K. {\em et al.}(2007)\cite{va}
consider the survival of the single life under multiple dependent
decrement causes,  their results are mainly  based on the copula
functions.  A statistical estimation on the net survival functions
using the observations of  crude survivals was proposed. Wang {\em
et al.}(2009) \cite{wa} has considered the decrement model for
comonotonic  group-life with status of joint-survival and last
survival under independent multiple decrement causes. The
distribution model with statistical analysis on comonotonic
group-life under multiple dependent decrement causes will be
considered in the present paper. The comonotonicity properties
seemingly make that each group-life can be represented by any
single member of the group, which would leads to a simple way
inducing a joint distribution for the considered collection of
comonotonic group-life. This paper is organized as follows.  In
Section 2, the main concepts like comonotonicity, copula
functions, survival function as well as some useful obtained
results are mentioned. in Section 3, we construct a model for
dependent multiple comonotonic group-life.

\section{Preliminaries}

\noindent\textbf{Definition 2.1} (Dhaene J.\cite{dh}) Let $R^n$ be a
$n$-dimensional Euclidean space, $A\subset R^n$ is said to be a
comonotonic vectors set if for any vectors
$\overline{x},\overline{y}\in A$, it holds that
$\overline{x}\leqslant\overline{y}$
 or $\overline{x}\geqslant\overline{y}$. Where $\overline{x}\leqslant(\geqslant)\overline{y}$
 means that $x_i \leqslant(\geqslant) y_i$, $i=1,\ldots,n$.

Following \cite{dh}, we see that the earlier concept of two
comonotonic random variables due to Schmeidler (1986), Yarri(1987)
's work on economics. Let $(X,Y)$ be a two dimensional random
vector on the probability space $(\Omega,\mathcal{F},P)$. If for
any $\omega_1,\omega_2\in\Omega$ there always holds
that$(X(\omega_1)-X(\omega_2))(Y(\omega_1)-Y(\omega_2))\geqslant
0$, then, $(X,Y)$ is said to be comonotonic,  this definition was
weaken  as
$(X(\omega_1)-X(\omega_2))(Y(\omega_1)-Y(\omega_2))\geqslant 0,
P.a.s.$ around 1997. For a $n$-dimensional random vector
$\overline{X}=(X_1,\ldots,X_n)$ defined on
$(\Omega,\mathcal{F},P)$, if there exists a subset $A\subset R^n$
such that $P(\overline{X}\in A)=1$, then $A$ is said to be a
support of $\overline{X}$.  A Random vector
$\overline{X}=(X_1,\ldots,X_n)$ is said to be comonotonic if its
support is comonotonic.

\noindent\textbf{Lemma 2.1}(Dhaene J. {\em et al.}\cite{dh}) A
random vector $\overline{X}=(X_1,\cdots,X_n)$ on probability space
$(\Omega,\mathcal{F},P)$ is comonotonic if and only if the
following equivalent conditions hold:\\
(1) $\overline{X}$ has a comonotonic support;\\
(2)For all values of $\overline{X}$, $\overline{x}=(x_1,\cdots,x_n)$, the joint distribution of $\overline{X}$ is
$$F_{\overline{X}}(\overline{x})=\min\{F_{X_1}(x_1),F_{X_2}(x_2),\ldots,F_{X_n}(x_n)\};$$
(3)Let $U$ be a random variable uniformly distributed on interval (0,1).  Then,
$$\overline{X}=^d (F_{X_1}^{-1}(U),F_{X_2}^{-1}(U),\ldots,F_{X_n}^{-1}(U));$$
(4)There exits random variable $z$ and non-decreasing real valued functions $f_i (i=1,\ldots,n)$ such that
$\overline{X}=^d(f_1 (z),f_2(z),\ldots,f_n(z)).$

Following the example of \cite{ca},\cite{ca1},\cite{va}, the
dependent multiple decrements  models for single life in life
insurance have been  investigated extensively. Here we only
introduce some basic notation and main proposed methods, for other
details the readers are referred to \cite{ca},\cite{ca1},\cite{va}
and the literatures therein.

We consider a group of  lives, exposed to $m$ competing causes of withdrawal from the group.
It is assumed that each individual may withdrawal from any single one of  the $m$ causes.
It is assumed that , at birth, each individual is assigned a vector of times
 $T_1,\cdots,T_m$, $0\leqslant T_j <\infty$, $j=1,\cdots,m$,
 representing individual's potential lifetime, if the individual were to withdrawal from each one of the $m$ causes.
 Obviously, the actual lifetime span is the minimum of all the $T_1,\cdots,T_m$.
 Thus, it is clear that under this model the lifetimes $T_1,\cdots,T_m$ are unobservable,
 and we can only observe the $\min(T_1,\cdots,T_m)$. In the classical multiple-decrement
 theory the random variables $T_1,\cdots,T_m$ are assumed independent, however, decrement
 in many real-life actuarial applications tend to be dependent, and the random variables
 $T_1,\cdots,T_m$ are considered stochastically dependent and also non-defective, i.e.
 $P(T_j < \infty)=1$. The joint distribution
$$
F(t_1,\cdots,t_m)=P(T_1\le t_1,\cdots,T_m\le t_m),
$$
the multivariate joint survival function
$$
S(t_1,\cdots,t_m)=P(T_1> t_1,\cdots,T_m> t_m)
$$
which is considered absolutely continuous and where $t_j\ge 0$, for $j=1,\cdots,m.$
The overall survival function of an individual aged $x\ge 0$
is defined through random variable $\min(T_1,\cdots,T_m)$ as
$$
\mathbb{S}(t):=S(t,\cdots,t)=P(T_1> t,\cdots,T_m>
t)=P(\min(T_1,\cdots,T_m)>t).
$$
The crude survival function $S^{(j)}(t)$ is defined as
$$S^{(j)}(t)=P(\min(T_1,\ldots,T_m)>t,\min(T_1,\cdots,T_m)=T_j).$$
The net survival function $S^{'(j)}(t)$ is defined as
$S^{'(j)}(t)=P(T_j >t)$. Note that $S^{'(j)}(t)$ is the marginal
survival function, due to cause alone, associated with the joint
multivariate survival function
$$
S(t_1,\cdots,t_m)=P(T_1> t_1,\cdots,T_m> t_m).
$$ Thus, we can view
$F(t_1,\cdots,t_m)=P(T_1\le t_1,\cdots,T_m\le t_m)$ as a
multivariate distribution with marginal distributions
$F^{'(j)}(t)=1-S^{'(j)}(t), j=1,\cdots,m.$ If we know
$S^{'(j)}(t)$, we can identify and calculate the joint survival
function $S(t_1,\cdots,t_m)$, and hence evaluate the overall
survival function $S(t,\cdots,t)$ under some assumption of copula
functions.

The copula is one of the most useful tools for handling
multivariate distributions with given univariate marginals
$F_1,\cdots,F_m$. Formally, a copula $C$ is a cumulative
distribution function, defined on $[0,1]^m$, with uniform
marginals. Given a copula $C$, if one defines
$$
F(x_1,\ldots,x_m)=C(F_1 (x_1),F_2 (x_2),\cdots,F_m (x_m)),
(x_1,\cdots,x_m)\in R^m,
$$
then $F$ is a multivariate distribution with univariate marginals
$F_1,\cdots,F_m.$  According to Sklar theorem (1959)\cite{va},
Given a continuous joint distribution function $F(x_1,\cdots,x_m)$
of $m$-dimensional random vector
 $\overline{X}=(X_1,\cdots,X_m)$, with marginal distribution functions
 $$
 F_1 (x_1),F_2 (x_2),\cdots,F_m (x_m), (x_1,\cdots,x_m)\in R^m,
 $$
 there corresponds to it a unique $m$-dimensional Copula function $C$
 can be constructed as
$$
C(u_1,\cdots,u_m)=F(F_1^{-1}(u_1),\cdots,F_m^{-1}(u_m)),
(u_1,\cdots,u_m)\in [0,1]^m.
$$
Note that different multivariate distributions $F$ may have the
same copula. Most of the multivariate dependence structure
properties of $F$ are in the copula function, which is independent
of the marginals and which is, in general, easier to handle than
the original $F$.

 The Sklar theorem can be restated to express the multivariate survival function $S(x_1,\cdots,x_n)$
 via an appropriate copula $\overline{C}$ called the survival copula of $(X_1,\cdots,X_n)$. Thus,
 $$
 S(x_1,\cdots,x_n)=\overline{C}(S_1(x_1),\cdots,S_n(x_n)).
 $$
Where $S_i(x_i)=1-F_i(x_i)=1-u_i,i=1,\cdots,n.$ When $n=2$, we
have $\overline{C}(1-u_1,1-u_2)=1-u_1-u_2+C(u_1,u_2).$  There are
many copula functions constructed by using the given
distributions, e.g. the multivariate Gaussian copula,  Student's
$t$-copula, and the popular Archimedean copulas  constructed by
$C^A(u_1,u_2)=\phi^{-1}(\phi(u_1)+\phi(u_2))$, where $\phi$ is a
continuous, convex function called generator, such that
$\phi(1)=0,\phi(0)=+\infty,$ of which the frequently used Clayton
copula, Gumbel copula,etc.(see. Nelsen (1999)\cite{ne}).

Some decrement models for independent as well as comonotonic
group-life with specified status under independent causes of
decrements have been obtained by Wang ( see. Wang Z.J. and Wang D.
(2009)\cite{wa}).

\section{Model for comonotonic group-life  under dependent decrement causes}
Let $x_1,\cdots,x_n$ be a group of $n$  lives with comonotonic
structure, and each member $x_i$ of the group can be withdrawn
with one of $m$ different dependent causes of decrements
$T_1,\cdots,T_m$. We start with the simple case where the status
of the lives group are determined as $g(x_1,\cdots,x_n)$, e.g.,
which can be the joint-survival status $(x_1\cdots x_n)$, or the
last-survival status $(\overline{x_1\cdots x_n})$,  etc..
$g(x_1,\cdots,x_n)$ can be withdrawn with anyone of $m$ different
dependent causes of decrements.
$$
T_1 (g(x_1,\cdots,x_n)),\cdots, T_m
(g(x_1,\cdots,x_n))
$$
are the latent decrement times of any status
$g(x_1,\cdots,x_n)$ of the group of $n$ lives. We can only observe
the time
$$
T(g(x_1,\cdots,x_n))=\min\{T_1(g(x_1,\cdots,x_n)),\cdots,T_m
(g(x_1,\cdots,x_n))\}.
$$
Since the dependency of the decrements causes, the latent
decrement times
$$
T_1 (g(x_1,\cdots,x_n)),\cdots, T_m (g(x_1,\cdots,x_n))
$$
are dependent. Assume that $T_1 (g(x_1,\cdots,x_n)),\cdots, T_m
(g(x_1,\cdots,x_n))$ has joint distribution function and marginals
as
$$
F_{T_1,T_2,\cdots,T_m} (t_1,t_2,\cdots,t_m)=P(T_1
(g(x_1,\cdots,x_n))\leqslant t_1, \cdots,T_m
(g(x_1,\cdots,x_n))\leqslant t_m);
$$
$$
F_{T_1}(t_1)=P(T_1 (g(x_1,\cdots,x_n))\leqslant t_1),\cdots,
F_{T_m}(t_m)=P(T_m (g(x_1,\cdots,x_n))\leqslant t_m),
$$
respectively. According to the Sklar theorem, there exists a
copula function $C$ such that
$$
F_{T_1,T_2,\cdots,T_m}
(t_1,t_2,\cdots,t_m)=C(F_{T_1}(t_1),\cdots,F_{T_m}(t_m)).
$$

\noindent\textbf{Example 1} Consider the case where $g(x,y)=(xy)$
and $g(x,y)=(\overline{xy})$, and lives $(x),(y)$ are comonotonic
and $m=2$.

(1) Let $F_{T_1, T_2}(t_1,t_2)$ be the joint distribution function
of $T_1 (xy),T_2(xy)$. Let $F_{T_1}(t_1),F_{T_2}(t_2)$ be the
distribution functions of $T_1 (xy),T_2(xy)$ respectively. Then,
there exists a Clayton copula function $C$ such that
$$
F_{T_1, T_2}(t_1,t_2)=\Big[(F_{T_1}(t_1))^{-\theta}+
(F_{T_2}(t_2))^{-\theta}-1\Big]^{-\frac{1}{\theta}},
$$
where the parameter $\theta>0,$ and
$$
F_{T_1}(t_1)=\max\{_{t_1}q_x^{(1)},_{t_1}q_y^{(1)}\},
F_{T_2}(t_2)=\max\{_{t_2}q_x^{(2)},_{t_2}q_y^{(2)}\}
$$
by the lemma 2.1 (3). Where $_tq^{(i)}_x:=P(T_i(x)\leqslant t)$
and so are in the sequel.

(2) The conditions of (1) but $g(x,y)=(\overline{xy})$. Then,
there exists Gumbel copula function $C$ such that
$$
F_{T_1, T_2}(t_1,t_2)=exp\Big\{-\Big[(-ln(F_{T_1}(t_1)))^{\theta}+
(-ln(F_{T_2}(t_2)))^{\theta}\Big]^{\frac{1}{\theta}}\Big\},
$$
where the parameter $\theta>0$ can be determined by the given
values of the Kandel statistics $\tau_\theta=1-\frac{1}{\theta}$,
and
$$
F_{T_1}(t_1)=\min\{_{t_1}q_x^{(1)},_{t_1}q_y^{(1)}\},
F_{T_2}(t_2)=\min\{_{t_2}q_x^{(2)},_{t_2}q_y^{(2)}\}
$$
by the lemma 2.1 (3).

(3) For the observable decrement time $T(g(x,y))$, $g(x,y)=(xy)$,
we have
\begin{eqnarray*}
_t q_{xy}^{(\tau)}&=& P(T(xy)\leqslant t,j=1,2)\\
&=& P(\min\{T_1 (xy),T_2 (xy)\}\leqslant t)\\
&=& 1-
\overline{C}\Big(1-\max\{_{t}q_x^{(1)},_{t}q_y^{(1)}\},1-\max\{_{t}q_x^{(2)},_{t}q_y^{(2)}\}\Big),
\end{eqnarray*}
where $\overline{C}$ is a survival copula function which can be
transformed into Gumbel or Clayton form. Similarly, we have
$$
_t q_{\overline{xy}}^{(\tau)}=1-
\overline{C}\Big(1-\min\{_{t}q_x^{(1)},_{t}q_y^{(1)}\},1-\min\{_{t}q_x^{(2)},_{t}q_y^{(2)}\}\Big)
$$
for status $g(x,y)=(\overline{xy})$.\\
Note that here the decrement probabilities $_t q_x^{(1)},_t
q_y^{(1)}, _t q_x^{(2)},_t q_y^{(2)}$, in fact, are the net
decrement probabilities corresponding to the net survival
functions $S^{'(1)} (t), S^{'(2)} (t)$, we could not obtain them
by the intuitive observations. According to Carriere JF.
(1994)\cite{ca} and V.K. Kaishev {\em et al.} (2007)\cite{va},
using a nonlinear system of differential equations which represent
the functional relation between the crude survival function
$S^{(j)}(t)$ and the net survival function $S^{'(j)}(t)$, under
condition that $S^{(j)}(t)$ can be estimated based on intuitive
observations, we can solve and estimate $S^{'(j)}(t)$.  Therefore,
we can obtain the estimates of $_t q_x^{(1)},_t q_y^{(1)}, _t
q_x^{(2)},_t q_y^{(2)}$.

Consider a general case where the status of the group of $n$ lives
are not specified, and the group of $n$ lives is comonotonic  and
each life of the group exposed to $m$ dependent decrements causes.
Then, we have $m$ dependent comonotonic future lifetime vectors as
$\overline{T}_i :=(T_i(x_1),\cdots, T_i(x_n)),i=1,\cdots,m.$ In
fact, the conclusion that there are dependent relations between
the  vectors $\overline{T}_1,\cdots,\overline{T}_m$ can be easily
verified by that the covariance matrix $COV (\overline{T}_k,
\overline{T}_j)\ne O,k\ne j$ (since $Cov (T_i(x_l),T_j(x_l))\ne
0.$) Note that this dependency in general can be modeled by the
multivariate distribution function of the vector
$(\overline{T}_1,\cdots,\overline{T}_m)$ which has a dimension
$nm$, we also note that each component vector $\overline{T}_i$ has
a joint distribution $F_{\overline{T}_i}(\overline{t}_i)$ which
equals to a comonotonic copula $\min\{F_{T_i
(x_1)}(t_{i1}),\cdots,F_{T_i (x_n)}(t_{in})\}$ by Lemma 2.1 (2),
where $\overline{t}_i:=(t_{i1},\cdots,t_{in})$. It is desirable to
construct the multivariate distribution function $F$ of the vector
$(\overline{T}_1,\cdots,\overline{T}_m)$ with the multivariate
marginals $F_{\overline{T}_1},\cdots,F_{\overline{T}_m}$, the
comonotonic copulas $\min\{F_{T_i (x_1)}(t_{i1}),\cdots,F_{T_i
(x_n)}(t_{in})\},i=1,\cdots,m$. According to H.Lin {\em et
al.}(1996) only the copula of independent case  can be considered
for modeling the joint multivariate distribution function with
the distributions of multivariate marginals, i.e. if
$$
F_{\overline{T}_1,\cdots,\overline{T}_m}
=C(F_{\overline{T}_1},\cdots F_{\overline{T}_m}).
$$
then the copula $C(u_1,\cdots,u_m)=\prod_{i=1}^m u_i.$ However,
this is not our case. The comonotonicity makes things much easier
than above case.

\noindent\textbf{Theorem 3.1} {\em Let $x_1,\cdots,x_n$ be a
comonotonic group-life exposed to $m$ dependent decrement causes
$T_1,\cdots,T_m$. Assume that for one member $x_l$ of the group
the future lifetime vector $(T_1(x_l),\cdots,T_m(x_l))$ has a
joint distribution $F_{T_1(x_l),\cdots, T_m(x_l)}$ with marginals
$F_{T_1(x_l)},\cdots,F_{T_m(x_l)}$. Then, their copula $C$ derived
from Sklar theorem is a copula for modelling the dependency in
comonotonic future lifetime vectors
$\overline{T}_1,\cdots,\overline{T}_m.$}

\noindent\textbf{\em Proof} By Sklar theorem there exists a copula
function $C$ such that
$$
F_{T_1(x_l),\cdots
,T_m(x_l)}(t_{1l},\cdots,t_{ml})=C(F_{T_1(x_l)}(t_{1l}),\cdots,F_{T_m(x_l)}(t_{ml})),
$$
and $F_{T_i(x_l)}(t_{il})\geqslant \min\{F_{T_i
(x_1)}(t_{i1}),\cdots,F_{T_i
(x_n)}(t_{in})\}=F_{\overline{T}_i}(\overline{t_i}),i=1,\cdots,m$,
since copula $C$ is a distribution function, thus we have
$$
C(F_{T_1(x_l)}(t_{1l}),\cdots,F_{T_m(x_l)}(t_{ml}))\geqslant
C(F_{\overline{T}_1}(\overline{t}_1),\cdots,F_{\overline{T}_m}(\overline{t}_m)),
$$
the most dependency structure properties of the vectors
$\overline{T}_1,\cdots,\overline{T}_m$ are in the later function
$C(F_{\overline{T}_1}(\overline{t}_1),\cdots,F_{\overline{T}_m}(\overline{t}_m))$,
which can be defined as a joint distribution function
$$
F(t_{11},\cdots,t_{1n},t_{21},\cdots,t_{2n},\cdots,t_{m1},\cdots,t_{mn})
$$
of $mn$-dimension. $\square$

\noindent\textbf{Corollary 3.1}  {\em Assume the condition of
Theorem 3.1. Then, the copula $C$ is also a survival copula for
the joint vector survival function
$S_{\overline{T}_1,\cdots,\overline{T}_m}(\overline{t}_1,\cdots,\overline{t}_m)$,
i.e.
$$
S_{\overline{T}_1,\cdots,\overline{T}_m}(\overline{t}_1,\cdots, \overline{t}_m)=
C(S_{\overline{T}_1 }(\overline{t}_1),\cdots, S_{\overline{T}_m} (\overline{t}_m)).
$$}
\noindent\textbf{\em Proof} By Lemma 2.1 (3), we have
\begin{eqnarray*}
S_{\overline{T}_i} (\overline{t}_i)&=&P(T_i(x_1)>t_{i1},\cdots,T_i(x_n)>t_{in})
=P(F^{-1}_{T_i(x_1)}(U)>t_{i1},\cdots, F^{-1}_{T_i(x_n)}(U)>t_{in})\\
&=& P(U>F_{T_i(x_1)}(t_{i1}), \cdots, U>F_{T_i(x_n)}(t_{in}))=
\max\{F_{T_i(x_1)}(t_{i1}),\cdots,F_{T_i(x_n)}(t_{in})\}\\
&\geqslant & F_{T_i(x_l)}(t_{il}), i=1,\cdots,m.
\end{eqnarray*}
Thus,
$$
C(F_{T_1(x_l)}(t_{1l}),\cdots,F_{T_m(x_l)}(t_{ml}))\leqslant C(S_{\overline{T}_1}
(\overline{t}_1),\cdots, S_{\overline{T}_m} (\overline{t}_m)),
$$
the later copula can be viewed as a survival copula and we define
it as the joint vector survival function
$S_{\overline{T}_1,\cdots,\overline{T}_m}(\overline{t}_1,\cdots,\overline{t}_m)$.
$\square$

The overall survival, crude survival for the future life time
vectors can be similarly defined. It is important to note that
$\min\{\overline{T}_1,\cdots,\overline{T}_m\}$ is the observable
decrement time of whole group-life. The crude survival
$S^{(j)}(\overline{t}):=P(\min\{\overline{T}_1,\cdots,\overline{T}_m\}>\overline{t},
\min\{\overline{T}_1,\cdots,\overline{T}_m\}=\overline{T}_j).$ If
we view the joint vector survival function
$S_{\overline{T}_1,\cdots,\overline{T}_m}(\overline{t}_1,\cdots,\overline{t}_m)$
as a $mn$-dimensional joint survival function
$$
S_{T_1(x_1),\cdots,T_1(x_n),\cdots,T_m(x_1),\cdots,T_m(x_n)}(t_{11},\cdots,t_{1n},\cdots,t_{m1},\cdots,t_{mn}),
$$
Then, the crude survival function for the case where $x_l$ is selected as a representative of the group can be defined as
$
S^{(j)}(t)=P(\min\{T_1(x_1),\cdots,T_1(x_n),\cdots,T_m(x_1),\cdots,T_m(x_n)\}>t,$\\
$\min\{T_1(x_1),\cdots,T_1(x_n),\cdots,T_m(x_1),\cdots,T_m(x_n)\}=T_j(x_l)).$

\noindent\textbf{Theorem 3.2}{\em (1) If the copula C in theorem 3.1 and corollary 3.1 is differentiable
with respect to $j$th variable and $S_{\overline{T}_j}(\overline{t}_j)$ is partially differentiable with
respect to $t_{jl}>0$ for all $j=1,\cdots,m,$ then
$$
\frac{d}{dt} S^{(j)}(t)=C_j (S_{\overline{T}_1} (\overline{t}_1),\cdots, S_{\overline{T}_m}
(\overline{t}_m))\times \frac{\partial (S_{\overline{T}_j} (\overline{t}_j))}{\partial t_{jl}}\Bigg\arrowvert_{t_{jl}=t}.
$$
where $C_j (u_1,\cdots,u_m)=\frac{\partial}{\partial
u_j}C(u_1,\cdots,u_m)$.\\
 (2)If the copula $C$ is $n$-order
differentiable with respect to its every variables and each
survival function $S_{\overline{T}_j}(\overline{t}_j)$ is
$n$-order partially differentiable with respect to its every
variables, then
$$
\frac{\partial^{(n)}}{\partial t_1\partial t_2\cdots\partial
t_n}S^{(j)}(\overline{t})=\frac{\partial^{(n)}}{\partial
t_{j1}\partial t_{j2}\cdots\partial t_{jn}} C(S_{\overline{T}_1}
(\overline{t}_1),\cdots, S_{\overline{T}_m}
(\overline{t}_m))\Bigg\arrowvert_{(t_{j1},\cdots,t_{jn})=(t_1,\cdots,t_n)},
$$
especially when $n=2$, it holds
$$
\frac{\partial^2 S^{(j)}(t_1,t_2)}{\partial t_1\partial
t_2}=\frac{\partial S_{\overline{T}_j}}{\partial
t_{j1}}\frac{\partial^2 C}{\partial
S_{\overline{T}_j}^2}\frac{\partial S_{\overline{T}_j}}{\partial
t_{j2}}\Bigg\arrowvert_{(t_{j1},t_{j2})=(t_1,t_2)}+ \frac{\partial
C}{\partial S_{\overline{T}_j}}\frac{\partial^2
S_{\overline{T}_j}}{\partial t_{j1}\partial
t_{j2}}\Bigg\arrowvert_{(t_{j1},t_{j2})=(t_1,t_2)},
$$
$j=1,\cdots,m.$}

\noindent\textbf{\em Proof} (1) See the proof of theorem 6 and
lemma 1 in Carriere JF.\cite{ca}(1994).\\
(2)Since
\begin{eqnarray*}
S^{(j)}(\overline{t})&=&P(\min\{\overline{T}_1,\cdots,\overline{T}_m\}>\overline{t},
\min\{\overline{T}_1,\cdots,\overline{T}_m\}=\overline{T}_j)\\
&=& P(\overline{T}_j>\overline{t}_j, \overline{T}_k
>\overline{T}_j (k\ne
j))=\int_{t_1}^\infty\cdots\int_{t_n}^\infty\Bigg\{\int_{t_{j1}}^\infty\cdots\int_{t_{jn}}^\infty\cdots
\int_{t_{j1}}^\infty\cdots\int_{t_{jn}}^\infty\\
& & f(t_{11},\cdots,t_{1n},\cdots,t_{m1},\cdots,t_{mn})\prod_{k\ne
j}dt_{k1}\cdots dt_{kn}\Bigg\}dt_{j1}\cdots dt_{jn}\\
&=&\int_{t_1}^\infty\cdots\int_{t_n}^\infty\Bigg\{
\frac{\partial^{(n)}}{\partial t_{j1}\partial t_{j2}\cdots\partial
t_{jn}}S_{T_1(x_1),\cdots,T_1(x_n),\cdots,T_m(x_1),\cdots,T_m(x_n)}\\
&
&(t_{11},\cdots,t_{1n},\cdots,t_{m1},\cdots,t_{mn})\Bigg\arrowvert_{(t_{k1,\cdots,t_{kn}})
=(t_{j1},\cdots,t_{jn}),\forall k} \Bigg\}dt_{j1}\cdots dt_{jn}\\
&=& \int_{t_1}^\infty\cdots\int_{t_n}^\infty\Bigg\{
\frac{\partial^{(n)}}{\partial t_{j1}\partial t_{j2}\cdots\partial
t_{jn}}C(S_{\overline{T}_1} (\overline{t}_1),\cdots,
S_{\overline{T}_m}
(\overline{t}_m))\Bigg\arrowvert_{(t_{k1,\cdots,t_{kn}})
=(t_{j1},\cdots,t_{jn}),\forall k} \Bigg\}dt_{j1}\cdots dt_{jn}.
\end{eqnarray*}
Thus,
$$
\frac{\partial^{(n)}}{\partial t_1\partial t_2\cdots\partial
t_n}S^{(j)}(\overline{t})=\frac{\partial^{(n)}}{\partial
t_{j1}\partial t_{j2}\cdots\partial t_{jn}} C(S_{\overline{T}_1}
(\overline{t}_1),\cdots, S_{\overline{T}_m}
(\overline{t}_m))\Bigg\arrowvert_{(t_{j1},\cdots,t_{jn})=(t_1,\cdots,t_n)}.
$$
$\square$

Theorem 3.2 indicates that in the case of comonotonic group-life,
the relationship between crude survivals and net survivals is
formulated by some much complicated $n$-order non-linear partial
differential equations, which in most case is unsolvable when we
have some knowledge of the crude survivals. When $n=2$, with the
conclusion (2) of theorem 3.2, If we have obtained the copula
function $C$ through considering some member of the group, then by
this fixed copula $C$, we may use some numerical methods
introduced in Carriere JF.\cite{ca} and Vladimir K Kaishev {\em et
al.}\cite{va} to obtain the estimates of
$S_{\overline{T}_1}(\overline{t}_1),\cdots,S_{\overline{T}_m}(\overline{t}_m)$
based on the observations of the crude survivals
$S^{(j)}(\overline{t}),j=1,\cdots,m$. Then the estimation of joint
survival
 $S_{\overline{T}_1,\cdots,\overline{T}_m}(\overline{t}_1,\cdots,\overline{t}_m)$ can be carried out.

\end{document}